\documentclass[prl, twocolumn, showpacs]{revtex4}
\usepackage{graphicx}
\usepackage{dcolumn}

\def\degC{${^\circ}$C}

\def\PRLsection #1 {{\it #1 --  }}  

\begin{document}

\title{Icosahedral Ti-Zr-Ni: A groundstate quasicrystal?}

\author{R. G. Hennig}
\altaffiliation[Current address: ]
{Ohio State University, Department of Physics, Columbus, OH 43210, USA.}
\author{A. E. Carlsson}
\author{K. F. Kelton}
\affiliation{Department of Physics, Washington University, St.\ Louis, MO
  63130, USA}
\author{C. L. Henley}
\affiliation{Laboratory of Atomic and Solid State Physics, Cornell University,
  Ithaca, NY 14853, USA}

\date{\today}

\begin{abstract}
  The first complete {\it ab initio} zero-temperature ternary phase
  diagram is constructed from the calculated energies of the
  elemental, binary and ternary Ti-Zr-Ni phases; for this, the
  icosahedral $i$-TiZrNi quasicrystal phase is approximated by
  periodic structures of up to 123 atoms/unit cell, based on a
  decorated-tiling model. The approximant structures containing the
  45-atom Bergman cluster were nearly degenerate, and stable against
  the competing binary phases. It is speculated that $i$-TiZrNi may be
  a ground state quasicrystal, as it is experimentally the
  low-temperature phase for its composition.
\end{abstract}


\maketitle

Thermodynamically stable, long-range ordered quasicrystals of
icosahedral symmetry are known in both of the main structural classes
of quasicrystal, the Al-transition metal class (e.g. $i$-AlPdMn) and
the Frank-Kasper class~\cite{Hen86a} (e.g. $i$-ZnMgY). The Ti-based
quasicrystals such as $i$-TiCrSiO~\cite{Hennig97} and the thermally
stable $i$-TiZrNi~\cite{Kelton97,Davis00,Hennig01} fall into either
respective class. They are of technical interest due to their high
melting point and capacity for hydrogen absorption, but are less
studied because the crystallites are small and the coherence length is
only 35~nm.

In experiments on the TiZrNi alloy, the crystal structure $W$-TiZrNi
is stable at high temperatures, but upon cooling it undergoes a
reversible phase transition at 570\degC~ to $i$-TiZrNi, showing that
$i$-TiZrNi is lower in energy than $W$-TiZrNi~\cite{Kelton97,Yi00}.
In fact $W$-TiZrNi is a ``periodic approximant'' of $i$-TiZrNi,
meaning that the unit cell is identical to a fragment of the
icosahedral. Long-time anneals (up to one month) at 500\degC~ gave no
indication that the quasicrystal transforms to some other phase. The
situation here contrasts with the Al-transition metal class, in which
the analogous crystal $\alpha$-AlMnSi (known as ``1/1 approximant''),
is lower in energy than the quasicrystal of identical composition.

In this work, we address the possibility that the $i$-TiZrNi
quasicrystal (or a very large unit cell crystal of nearly identical
structure) is in fact a ground state. We compute the ternary ground
state phase diagram, using a database of {\it ab initio} total
energies of 28 elemental, binary and ternary crystalline phases in the
Ti-Zr-Ni alloy system.  The total energy for the quasicrystal phase is
determined from six different periodic approximants of $i$-TiZrNi.
This calculation depends on our previously reported structure
model~\cite{Hennig01}, which was formulated as a decoration by atoms
of the canonical cell tilings, and fitted to a combination of
diffraction data and {\em ab initio} relaxations.

\PRLsection{Ti-Zr-Ni crystalline phase diagram} To validate the
experimental findings as well as the diffraction-based structure
model~\cite{Hennig01}, we computed the energies of quasicrystal-like
and competing structures and from this constructed the complete ground
state ternary phase diagram. This is the first {\it ab initio}
investigation of the Ti-Zr-Ni alloy system.

Our {\it ab initio} total energy calculations were performed with
VASP~\cite{VASP}, which is a density functional code using a
plane-wave basis and ultrasoft Vanderbilt type
pseudopotentials~\cite{VanderbiltPP}. The calculations were performed
using the generalized gradient approximation by Perdew and
Wang~\cite{Perdew92a} and a plane-wave kinetic-energy cutoff of
$E_\mathrm{cut} = 302.0\,$eV was chosen to ensure convergence of the
energy. The pseudopotentials for Ti and Zr describe the 3p and 4p
states, respectively, as semi-core states. This was found to be
necessary to avoid unphysical short distances between Ti and Zr atoms.
Atomic-level forces were calculated and relaxations with a conjugate
gradient method were performed. The positions of the atoms, as well as
the shape and volume of the unit cells, were relaxed until the total
electronic energy changed by less than $1\,$meV. This corresponds to
atomic-level forces $F_\mathrm{max} \leq 0.02\,$eV/\AA.  The size of
the k-point mesh was chosen to give the same accuracy for the energy.

The ground state phase diagram is constructed by calculating the
ground state energy surface. This corresponds to determining the
convex hull of the set of energy points as a function of the
composition, as determined by the energy calculations for the
different phases. Before investigating the whole ternary ground state
phase diagram, the phase diagram for the binary phases is determined
from the energy calculations. For each concentration the lowest energy
structure is determined (see Table~\ref{tab:Energies}). The convex
hull of these points is then constructed by considering chemical
equilibria between the different phases.

\begin{table}[htbp]
  \caption {Structures and electronic energies of the competing
    {\rm Ti-Zr-Ni} phases: The calculated energies, $E$, the energy difference to the
    groundstate, $\Delta E$, and the heats of formation, $\Delta
    H_\mathrm{f}$, are given. Wherever available, experimental values are
    given in parentheses~\cite{Boer88}.}
  \label{tab:Energies}
  \begin{ruledtabular}
    \begin{tabular}{l c c c c r@{~~} r l}
      Structure & \multicolumn{3}{c}{$N_\mathrm{atoms}$} & $E$ &
      \multicolumn{1}{c}{$\Delta E$} &
      \multicolumn{2}{c}{$\Delta H_\mathrm{f}$} \\
      & Ti & Zr & Ni & $\left [\frac{\mathrm{eV}}{\mathrm{atom}} \right ]$ &
      \multicolumn{1}{c}{$\left [\frac{\mathrm{meV}}{\mathrm{atom}} \right]$} &
      \multicolumn{2}{c}{$\left [\frac{\mathrm{kJ}}{\mathrm{mol}} \right]$} \\
      \colrule
      $\alpha$-Ti\footnotemark[1] (A3)    &  2 &  0 &  0 & $-7.752$ &    0 & \multicolumn{2}{c}{--} \\
      $\beta$-Ti\footnotemark[2]  (A2)    &  1 &  0 &  0 & $-7.647$ & +105 & \multicolumn{2}{c}{--} \\
      $\gamma$-Ti (A1)                    &  1 &  0 &  0 & $-7.704$ &  +48 & \multicolumn{2}{c}{--} \\
      $\alpha$-Zr\footnotemark[1] (A3)    &  0 &  2 &  0 & $-8.398$ &    0 & \multicolumn{2}{c}{--} \\
      $\beta$-Zr\footnotemark[2]  (A2)    &  0 &  1 &  0 & $-8.351$ &  +47 & \multicolumn{2}{c}{--} \\
      $\gamma$-Zr (A1)                    &  0 &  1 &  0 & $-8.344$ &  +54 & \multicolumn{2}{c}{--} \\
      $\gamma$-Ni\footnotemark[1] (A1)    &  0 &  1 &  0 & $-5.422$ &    0 & \multicolumn{2}{c}{--} \\
      $\alpha$-Ni (A3)                    &  0 &  2 &  0 & $-5.410$ &  +12 & \multicolumn{2}{c}{--} \\
      $\beta$-Ni  (A2)                    &  0 &  1 &  0 & $-5.380$ &  +42 & \multicolumn{2}{c}{--} \\
      \colrule
      Ti$_2$Ni\footnotemark[1] (E9$_3$)   & 16 &  0 &  8 & $-7.286$ &    0 & $-30$ &($-27$) \\
      Ti$_2$Ni (C16)                      &  4 &  0 &  2 & $-7.278$ &   +7 & $-29$ \\
      TiNi\footnotemark[1] (B$19'$)       &  2 &  0 &  2 & $-7.037$ &   +3 & $-43$ &($-34$) \\
      TiNi\footnotemark[2] (B2)           &  1 &  0 &  1 & $-6.989$ &  +49 & $-39$ \\
      TiNi (B$_\mathrm{f}$)               &  2 &  0 &  2 & $-7.040$ &    0 & $-44$ \\
      TiNi$_3$\footnotemark[1] (D0$_{24}$)&  4 &  0 & 12 & $-6.546$ &    0 & $-52$ &($-35$) \\
      \colrule
      Zr$_2$Ni\footnotemark[1] (C16)      &  0 &  4 &  2 & $-7.759$ &    0 & $-34$ &($-37$) \\
      Zr$_2$Ni (E9$_3$)                   &  0 & 16 &  8 & $-7.665$ &  +93 & $-25$ \\
      ZrNi\footnotemark[1] (B$_\mathrm{f}$)& 0 &  2 &  2 & $-7.407$ &    0 & $-48$ &($-49$) \\
      ZrNi (B2)                           &  0 &  1 &  1 & $-7.298$ & +109 & $-37$ \\
      ZrNi (B$19'$)                       &  0 &  2 &  2 & $-7.407$ &   +1 & $-48$ \\ 
      Zr$_7$Ni$_{10}$\footnotemark[1] ($Aba2$)&0& 14& 20 & $-7.143$ &   +5 & $-50$ &($-52$) \\
      Zr$_7$Ni$_{10}$\footnotemark[1] ($Pbca$)&0& 28& 40 & $-7.143$ &   +5 & $-50$ &($-52$) \\
      ZrNi$_2$ (C15)                      &  0 &  2 &  4 & $-6.864$ &  +54 & $-43$ &($-73$) \\
      ZrNi$_3$\footnotemark[1] (D0$_{19}$)&  0 &  2 &  6 & $-6.674$ &    0 & $-49$ &($-67$) \\ 
      Zr$_2$Ni$_7$\footnotemark[1] ($C_2/m$)&0 &  4 & 14 & $-6.559$ &    0 & $-46$ &($-46$) \\
      ZrNi$_5$\footnotemark[1] (C15$_\mathrm{b}$)&0&1& 5 & $-6.287$ &    0 & $-36$ &($-35$) \\
      \colrule
      $\delta$-Ti$_6$Zr$_2$Ni$_4$ (E$9_3$)& 12 &  4 &  8 & $-7.335$ &  +69 & $-24$ \\
      $\delta$-Ti$_2$Zr$_6$Ni$_4$ (E$9_3$)&  4 & 12 &  8 & $-7.599$ &  +41 & $-29$ \\
      $\lambda$-Ti$_4$Zr$_4$Ni$_4$\footnotemark[2] (C14)&4&4&4&$-7.387$&+135&$-19$ \\
      $\lambda$-Ti$_6$Zr$_4$Ni$_2$ (C14)  &  6 &  4 &  2 & $-7.767$ &$-11$ & $-18$ & 
      \footnotetext[1]{Experimentally observed groundstate phase.}
      \footnotetext[2]{Experimentally observed high-temperature phase.}
    \end{tabular}
  \end{ruledtabular}
\end{table}

\begin{figure}[htb]
 \includegraphics[width=8.5cm] {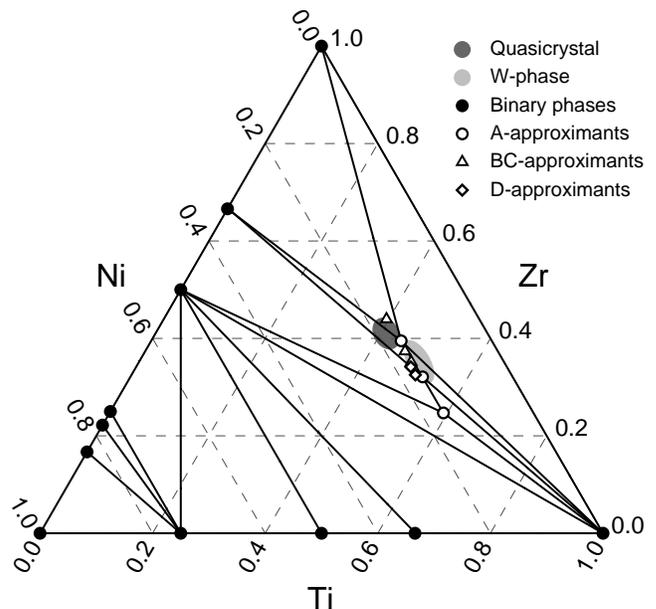} \vspace{-0.5cm}
  \caption {Ground state phase diagram of\/ {\rm
      Ti-Zr-Ni} showing the creases (solid lines) of the minimum
    energy surface of the competing phases. Each crease represents a
    domain of coexistence of the two phases at that crease's end
    points. Each three-phase coexistence domain is a triangle bounded
    by creases with endpoints at the corresponding single-phase
    points. Shaded regions represent the experimental composition of
    the quasicrystal and the approximant phase.}
  \label{fig:Phasediagram}
\end{figure}

For the Ti-Ni phase diagram, we find in agreement with the
experimental phase diagram~\cite{Massalski86} the ground state phases
$\alpha$-Ti (hcp), $\gamma$-Ni (fcc), $\delta$-Ti$_2$Ni (E9$_3$), and
TiNi$_3$ (D0$_{24}$). The TiNi martensite structure
B$19'$~\cite{Marcinkowski68} is found to be lower in energy than the
high-temperature phase B2 and nearly degenerate with the
B$_\mathrm{f}$ structure, which to our knowledge has not been
observed.

The ground state phases in the computed Zr-Ni phase diagram are, in
addition to $\alpha$-Zr (hcp) and $\gamma$-Ni (fcc), the binary phases
Zr$_2$Ni (C16), ZrNi (B$_\mathrm{f}$), ZrNi$_3$ (D0$_{19}$) and
ZrNi$_5$ (C15$_\mathrm{b}$). This is consistent with the binary phase
diagram~\cite{Massalski86}. The calculated energies for both
Zr$_7$Ni$_{10}$ phases, $Pbca$ and $Aba2$, indicate that these phases
are slightly unstable ($\Delta E = 5\,$meV/atom) and will decompose
into ZrNi and ZrNi$_3$.  Our {\it ab initio} calculation of the binary
ground state phase diagram and of the heats of formation
(Table~\ref{tab:Energies}) show excellent agreement with experiment.
Furthermore, our calculated lattice parameters (not shown here) are
within 2\% of the experimental values for all observed structures.

Besides the approximant phase $W$-TiZrNi and the quasicrystal
$i$-TiZrNi, two ternary TiZrNi phases are observed for Ni
concentrations below 50\%, the hexagonal $\lambda$-TiZrNi
phase~\cite{Molokanov89, Majzoub01} and the cubic
$\delta$-(Ti,Zr)$_2$Ni phase~\cite{Yurko59, Mackay94}. The $\delta$
phase is stabilized by small amounts of oxygen and is not found in
samples low in oxygen~\cite{Mackay94}; our calculations
(Table~\ref{tab:Energies}) found it to be unstable with respect to the
binary phases. The atomically disordered C14 Laves phase
$\lambda$-TiZrNi is often seen in quasicrystal samples, but is stable
only at high temperatures~\cite{Davis00}. Total energy calculations
for nine deterministic ternary variants of the C14 phase showed all to
be unstable against decomposition into binary phases, except for the
nickel-poor $\lambda$-Ti$_6$Zr$_4$Ni$_2$; but that in turn is computed
to be less stable than the $W$-TiZrNi ``approximant'' structure, which
has practically the same composition~\cite{Majzoub01}.

\PRLsection{Quasicrystal phase stability}
It is nontrivial to set up a computation of the quasicrystal total
energy, since there is no finite unit cell.  We take advantage of the
description of the quasicrystal as a space-filling tiling of a few
types of cell, each type containing a fixed placement (``decoration'')
by atoms. A {\it periodic} packing of the same tiles, after
decoration, forms an ``approximant'' structure.  To estimate the total
energy of the quasicrystal, we calculated that of several
approximants, each made from essentially one tile type.

The first, simpler tiling model~\cite{Hen86a} uses the ``Ammann''
cells, a prolate ($PR$) and an oblate ($OR$) rhombohedron with edges
of length 0.516~nm, as well as a composite tile called the rhombic
dodecahedron ($RD$). This model's decoration is Ni on vertices, Ti on
edges, and Zr (in the role of a larger atom) in interiors. Our
Ammann-cell approximants do not contain the Bergman cluster and have 4
to 19 atoms per periodic unit cell.

The second tiling model~\cite{Hen91a} uses larger ``canonical cells''
known as $A$, $B$, $C$, and $D$. Canonical cell structures can be
viewed as a particular way to group Ammann tiles, so as to maximize
the frequency of the icosahedrally symmetric 45-atom ``Bergman
cluster''. The cluster centers are the canonical cell tile corners,
linked by edges of 1.24~nm. This model's atomic decoration is locally
similar to the first model's, but permits many more site types
depending on the local context. In a previous paper~\cite{Hennig01},
we determined the structure of $i$-TiZrNi by a constrained
least-squares fit of this model, in which the chemical site
occupations were refined using X-ray and neutron diffraction data,
while the atomic positions were optimized by {\it ab initio}
relaxations of the resulting atomic decoration for periodic tilings;
two iterations of this two-step procedure gave convergence. Several
sites were found to have mixed occupation, which must be chosen one
way or the other in the {\it ab initio} calculations, producing the
numbered variant decorations in Table~\ref{tab:Energies-QC}. The
canonical-cell approximants ($A_6$, $B_2C_2$ and $D_2$ packings) are
relatively large, with 81 to 123 atoms per crystallographic unit
cell~\cite{Hen91a,Hennig01}.

\begin{table}[tb]
  \caption {Electronic energies of periodic tiling structures approximating
    $i$-TiZrNi. Energies, $E$, energy differences to the groundstate,
    $\Delta E$, and to the binary phases, $\Delta
    E_{\mathrm{binary}}$, are given.}
  \label{tab:Energies-QC}
  \begin{ruledtabular}
    \begin{tabular}{l c c c c c c}
      Structure & \multicolumn{3}{c}{$N_\mathrm{atoms}$} & $E$ &
      $\Delta E$ & $\Delta E_{\mathrm{binary}}$ \\
      & Ti & Zr & Ni & $\left [\frac{\mathrm{eV}}{\mathrm{atom}} \right ]$ & 
      $\left [\frac{\mathrm{meV}}{\mathrm{atom}} \right ]$  &
      $\left [\frac{\mathrm{meV}}{\mathrm{atom}} \right ]$ \\
      \colrule
      $OR$         &  3 &  0 &  1 & $-7.078$ & $+324$ &$+324$ \\
      $PR$         &  3 &  2 &  1 & $-7.764$ &  $+15$ &  $+3$ \\
      $RD$         &  8 &  8 &  3 & $-7.737$ & $+110$ & $+84$ \\
      $A_6$ (1)    & 36 & 32 & 13 & $-7.827$ &    $0$ & $-24$ \\
      $A_6$ (2)    & 42 & 26 & 13 & $-7.779$ &    $0$ & $-24$ \\
      $A_6$ (3)    & 42 & 26 & 13 & $-7.731$ &    $0$ & $-28$ \\
      $B_2C_2$ (1) & 42 & 34 & 15 & $-7.803$ &   $+4$ & $-19$ \\
      $B_2C_2$ (2) & 36 & 40 & 15 & $-7.839$ &   $+9$ & $-11$ \\
      $B_2C_2$ (3) & 44 & 32 & 15 & $-7.789$ &   $+4$ & $-19$ \\
      $D_2$ (1)    & 60 & 42 & 21 & $-7.773$ &   $+5$ & $-17$ \\
      $D_2$ (2)    & 62 & 40 & 21 & $-7.758$ &   $+9$ & $-14$ \\
      \colrule
      i-TiZrNi (1) &46.0\%&37.6\%&16.4\%& $-7.806$ & $+3$ & $-21$ \\
      i-TiZrNi (2) &42.7\%&40.9\%&16.4\%& $-7.824$ & $+6$ & $-17$
    \end{tabular}
  \end{ruledtabular}
\end{table}

From the energies of the ternary approximant structures the
groundstate energy surface is constructed in the same way as done for
the binary phases. Table~\ref{tab:Energies-QC} gives the energy of
formation $\Delta E$ we found for these structures, which is defined
as the energy difference to the coexisting mixture of three competing
binary phases with the same composition and the lowest possible
energy, i.e. the groundstate energy surface. $\Delta
E_\mathrm{binary}$ is defined in comparison to a coexisting mixture of
three competing binary phases, which in our calculation are
$\alpha$-Ti, Zr$_2$Ni, and (depending on the composition) either
$\alpha$-Zr or ZrNi. Experimentally, the competing phases of the
quasicrystal are $\alpha$-Ti/Zr, Ti$_2$Ni, Zr$_2$Ni, $\lambda$-TiZrNi
(Laves phase), and the $W$-TiZrNi approximant phase~\cite{Davis00}.

All three periodic Ammann tiling structures are unstable against the
competing binary phases. The $OR$ packing is particularly unfavorable
energetically. On the other hand, the $PR$ packing has $\Delta E
\approx 0$. This structure is equivalent to the cubic C15
(MgCu$_2$-type) Laves phase~\cite{Hen86a}; it can be obtained by
replacing most of the Ni atoms in the C15 ZrNi$_2$ structure by Ti.

The quasicrystal-like periodic canonical cell tilings, on the other
hand, have significantly lower energies, by up to $28\,$meV/atom
compared to the competing binary phases and also lower in energy than
the ternary C14 phase, $\lambda$-Ti$_6$Zr$_4$Ni$_2$, as mentioned
earlier. We found the three $A_6$ structures to be groundstate
structures and the $B_2C_2$ as well as the $D_2$ structures to lie
slightly above the groundstate energy surface by 4 to 9 meV/atom. The
$A_6$ tiling corresponds to the experimentally observed $W$-TiZrNi
phase (which inspired the structure model~\cite{Hennig01}).
Diffraction experiments show that $W$-TiZrNi has Ti/Zr disorder on the
two kinds of site which lie on the plane bisecting the $\langle
100\rangle$ axis linking two Bergman clusters~\cite{Hennig00};
therefore we investigated variant chemical occupations of these
so-called ``glue'' sites. We found that all three variants give the
lowest energy for their composition.  In fact, the three structures
form a line on the groundstate energy surface, indicating that Ti/Zr
disorder on these sites costs no energy.

The next larger canonical cell tilings, the $B_2C_2$ and the $D_2$
tilings, do not correspond to experimentally observed phases, but are
computed to be stable with respect to the binary phases. We
investigated variant site occupations in both these
structures~\cite{HennigThesis}. Changing Zr atoms to Ti on certain
sites only weakly influences the energy, whereas changing Ti atoms to
Ni leads to significant increase of the energy, indicating that the
quasicrystal might contain Ti/Zr disorder but not Ti/Ni disorder, in
agreement with the structural refinement of $i$-TiZrNi of
Ref.~\onlinecite{Hennig01}.

The stability of the decoration model of the three periodic canonical
cell tilings is a strong indication that the decoration of a larger
tiling should be practically stable. We estimated the energy of
$i$-TiZrNi by summing the energy of each constituent tile, as found
from the structures with only one kind of tile. In an infinite
icosahedral canonical-cell tiling, the number ratio of tiles is
$N(A):N(BC):N(D) = (3[3-\sqrt5] : 1 : [\sqrt5-2])$, where we adopted
the ``magic'' value $\zeta= 3(1-2/\sqrt5)$, a parameter for the
frequency of $D$ cells~\cite{Hen91a}. Taking the lowest energy tiling
structures $A_6$(1), $B_2C_2$(1) and the $D_2$(1) yields for the
composition of the quasicrystal Ti$_{46.0}$Zr$_{37.6}$Ni$_{16.4}$ (1),
4\% higher in Ti than the experimental composition of
Ti$_{41.5}$Zr$_{41.5}$Ni$_{17}$. Taking the Zr rich $B_2C_2$ (2)
structure instead yields Ti$_{42.7}$Zr$_{40.9}$Ni$_{16.4}$ (2) within
1\% of the experimental composition. The energies of the periodic
tilings are not enough to determine the energy of the quasicrystal
since inter-tile interactions (called ``tile Hamiltonian'') need to be
considered too~\cite{Mih96}. Neglecting any tile-tile interactions the
two quasicrystal models are 21 and 17~meV/atom lower than the
competing binary phases. Both structures, however, are slightly above
the ground state energy surface by 3 and 6~meV/atom
respectively~\footnote{ The (small) energy differences here are
  roughly half those calculated for model $i$-AlMn quasicrystals using
  {\it ab-initio} based pair potentials (See Ref.~\onlinecite{Mih96},
  Fig.~5.)}. This combined with the experimental results suggests that
the icosahedral TiZrNi quasicrystal really may be a ground state
quasicrystal.

What mechanism stabilizes the quasicrystal-like approximant
structures? Ti and Zr exhibit a zero heat of mixing. Ti and Ni as well
as Zr and Ni, on the other hand, show a strongly attractive
interaction. Thus, it is energetically favorable for the Ni atoms to
be surrounded by Ti or Zr. This is reflected in a large charge
transfer from the Ti/Zr to Ni and a strong hybridization between the Ni
and the Ti/Zr subbands which we observe in our calculations.
This explains why in the quasicrystalline structure the Ni atoms
occupy the sites along the edges of the Ammann tiles surrounded by Ti
and Zr. Furthermore, since Zr is slightly larger than Ti it is no
surprise that Zr occupies the more open sites of the structure.

However, the unstable small-Ammann-tiling approximants $PR$, $OR$, and
$RD$ have virtually the same {\it local} atomic structure as the
canonical-cell approximants $A_6$, $B_2C_2$, and $D_2$. The main
difference is that complete Bergman clusters are absent in the small
Ammann approximants, but dense in the canonical cell approximants
(where they contain $\sim 1/2$ of the atoms); we conjecture that this
is responsible for the energy difference.

\PRLsection{Conclusion} The first systematic investigation of the
energies of structures of the binary Ti-Ni and Zr-Ni phase diagrams
was presented and it was shown that {\it ab initio} calculations for
these systems yield excellent agreement for the ground state phases
with experiments. The phase transition between the $W$ phase and the
quasicrystal observed experimentally as well as the long-time
annealing experiments demonstrate that the quasicrystal is a stable
low temperature phase. The {\it ab initio} energies of the nickel-poor
Ti-Zr-Ni ternary crystalline phases showed that only the large-cell
``quasicrystal approximants'' containing complete Bergman clusters
were stable against decomposition into the binary phases. This
demonstrates that the $i$-TiZrNi phase, the $W$-TiZrNi phase, or a
similar large-cell approximant is the groundstate of the system. The
calculations (Table~\ref{tab:Energies-QC}) imply a {\it total} energy
difference favoring $W$-TiZrNi by the order of 100 meV per {\it tile}.
The experimental situation indicates that $i$-TiZrNi is stabilized by
energy. Conceivably the quasicrystal -- or large approximants that mix
different canonical cell types -- could be stabilized by the cell-cell
interaction energies, which were hitherto neglected. Energy
calculations on those larger approximants would permit the extraction
of the tile Hamiltonian, which is needed to decide whether the
quasicrystal is a ground state, and to address its long-range
structure.

\begin{acknowledgments}
  The work at Washington University was supported by the National
  Science Foundation (NSF) under grants DMR 97-05202 and DMR 00-72787.
  C.L.H. was supported by D.O.E. grant DE-FG02-89ER-45405. R.G.H.\ was
  partially supported by NSF grant DMR-0080766 and by DOE grant
  DE-FG02-99ER45795.
\end{acknowledgments}


\begin{thebibliography}{20}
\expandafter\ifx\csname natexlab\endcsname\relax\def\natexlab#1{#1}\fi
\expandafter\ifx\csname bibnamefont\endcsname\relax
  \def\bibnamefont#1{#1}\fi
\expandafter\ifx\csname bibfnamefont\endcsname\relax
  \def\bibfnamefont#1{#1}\fi
\expandafter\ifx\csname citenamefont\endcsname\relax
  \def\citenamefont#1{#1}\fi
\expandafter\ifx\csname url\endcsname\relax
  \def\url#1{\texttt{#1}}\fi
\expandafter\ifx\csname urlprefix\endcsname\relax\def\urlprefix{URL }\fi
\providecommand{\bibinfo}[2]{#2}
\providecommand{\eprint}[2][]{\url{#2}}

\bibitem{Hen86a}
\bibinfo{author}{\bibfnamefont{C.~L.} \bibnamefont{Henley}} \bibnamefont{and}
  \bibinfo{author}{\bibfnamefont{V.}~\bibnamefont{Elser}},
  \bibinfo{journal}{Phil.\ Mag.\ B}
  \textbf{\bibinfo{volume}{53}}, \bibinfo{pages}{L59} (\bibinfo{year}{1986}).

\bibitem{Hennig97}
  \bibinfo{author}{\bibfnamefont{R.~G.} \bibnamefont{Hennig}}
  \bibnamefont{and} \bibinfo{author}{\bibfnamefont{H.}~\bibnamefont{Teichler}},
  \bibinfo{journal}{Phil.\ Mag.\ A} \textbf{\bibinfo{volume}{76}},
  \bibinfo{pages}{1053} (\bibinfo{year}{1997}).

\bibitem{Kelton97}
  \bibinfo{author}{\bibfnamefont{K.~F.} \bibnamefont{Kelton}},
  \bibinfo{author}{\bibfnamefont{W.~J.} \bibnamefont{Kim}},
  \bibnamefont{and} \bibinfo{author}{\bibfnamefont{R.~M.} \bibnamefont{Stroud}},
  \bibinfo{journal}{Appl.\ Phys.\ Lett.} \textbf{\bibinfo{volume}{70}}
  (\bibinfo{year}{1997}).

\bibitem{Davis00}
  \bibinfo{author}{\bibfnamefont{J.~P.} \bibnamefont{Davis}},
  \bibinfo{author}{\bibfnamefont{E.~H.} \bibnamefont{Majzoub}},
  \bibinfo{author}{\bibfnamefont{J.~M.} \bibnamefont{Simmons}},
  \bibnamefont{and} \bibinfo{author}{\bibfnamefont{K.~F.}
    \bibnamefont{Kelton}}, \bibinfo{journal}{Mat.\ Sci.\ Eng.\ A}
  \textbf{\bibinfo{volume}{294-296}}, \bibinfo{pages}{104}
  (\bibinfo{year}{2000}).

\bibitem{Hennig01}
  \bibinfo{author}{\bibfnamefont{R.~G.} \bibnamefont{Hennig}},
  \bibinfo{author}{\bibfnamefont{K.~F.} \bibnamefont{Kelton}},
  \bibinfo{author}{\bibfnamefont{A.~E.} \bibnamefont{Carlsson}},
  \bibnamefont{and} \bibinfo{author}{\bibfnamefont{C.~L.}
  \bibnamefont{Henley}}, \bibinfo{journal}{Submitted to Phys.\ Rev.\ B,
  cond-mat/0202536} (\bibinfo{year}{2001}).

\bibitem{Yi00}
  \bibinfo{author}{\bibfnamefont{S.} \bibnamefont{Yi}}
  \bibnamefont{and} \bibinfo{author}{\bibfnamefont{D.~H.} \bibnamefont{Kim}},
  \bibinfo{journal}{J.\ Mat.\ Res.} \textbf{\bibinfo{volume}{15}}
  (\bibinfo{year}{2000}).

\bibitem{VASP}
  \bibinfo{author}{\bibfnamefont{G.}~\bibnamefont{Kresse}} \bibnamefont{and}
  \bibinfo{author}{\bibfnamefont{J.}~\bibnamefont{Hafner}},
  \bibinfo{journal}{Phys.\ Rev.\ B}
  \textbf{\bibinfo{volume}{47}}, \bibinfo{pages}{RC558} (\bibinfo{year}{1993}),
  \bibinfo{author}{\bibfnamefont{G.}~\bibnamefont{Kresse}} \bibnamefont{and}
  \bibinfo{author}{\bibfnamefont{J.}~\bibnamefont{Furthm\"uller}},
  \bibinfo{journal}{Comp.\ Mat.\ Sci.}
  \textbf{\bibinfo{volume}{6}}, \bibinfo{pages}{15} (\bibinfo{year}{1996}{\natexlab{a}}),
  \bibinfo{journal}{Phys.\ Rev.\ B}
  \textbf{\bibinfo{volume}{54}}, \bibinfo{pages}{11169} (\bibinfo{year}{1996}{\natexlab{b}}).

\bibitem{VanderbiltPP}
  \bibinfo{author}{\bibfnamefont{D.}~\bibnamefont{Vanderbilt}},
  \bibinfo{journal}{Phys.\ Rev.\ B}
  \textbf{\bibinfo{volume}{41}}, \bibinfo{pages}{7892} (\bibinfo{year}{1990}),
  \bibinfo{author}{\bibfnamefont{G.}~\bibnamefont{Kresse}} \bibnamefont{and}
  \bibinfo{author}{\bibfnamefont{J.}~\bibnamefont{Hafner}},
  \bibinfo{journal}{J.\ Phys.: Condens.\ Matter}
  \textbf{\bibinfo{volume}{6}}, \bibinfo{pages}{8245} (\bibinfo{year}{1994}).

\bibitem{Perdew92a}
  \bibinfo{author}{\bibfnamefont{J.~P.}~\bibnamefont{Perdew}} \bibnamefont{and}
  \bibinfo{author}{\bibfnamefont{Y.}~\bibnamefont{Wang}},
  \bibinfo{journal}{Phys.\ Rev.\ B}
  \textbf{\bibinfo{volume}{45}}, \bibinfo{pages}{13244} (\bibinfo{year}{1992}).
  
\bibitem{Boer88}
  \bibinfo{author}{\bibfnamefont{F.~R.}  \bibnamefont{{de Boer}}},
  \bibinfo{author}{\bibfnamefont{R.}~\bibnamefont{Boom}},
  \bibinfo{author}{\bibfnamefont{W.~C.~M.} \bibnamefont{Mattens}},
  \bibinfo{author}{\bibfnamefont{A.~R.} \bibnamefont{Miedema}},
  \bibnamefont{and} \bibinfo{author}{\bibfnamefont{A.~K.}
    \bibnamefont{Niessen}}, \emph{\bibinfo{title}{Cohesion in Metals:
      Transition Metal Alloys}}
  (\bibinfo{publisher}{Amsterdam: North-Holland},
  \bibinfo{year}{1988}).

\bibitem{Massalski86}
  \bibinfo{author}{\bibfnamefont{T.~B.} \bibnamefont{Massalski}},
  \bibinfo{author}{\bibfnamefont{J.~L.} \bibnamefont{Murray}},
  \bibinfo{author}{\bibfnamefont{L.~H.} \bibnamefont{Bennett}},
  \bibnamefont{and} \bibinfo{author}{\bibfnamefont{H.}~\bibnamefont{Baker}},
  \emph{\bibinfo{title}{Binary Alloy Phase Diagrams}}
  (\bibinfo{publisher}{Metals Park, Ohio: American Society for Metals},
  \bibinfo{year}{1986}).

\bibitem{Marcinkowski68}
  \bibinfo{author}{\bibfnamefont{M.~J.} \bibnamefont{Marcinkowski}},
  \bibinfo{author}{\bibfnamefont{A.~S.} \bibnamefont{Sastri}},
  \bibnamefont{and} \bibinfo{author}{\bibfnamefont{D.} \bibnamefont{Koskimaki}},
  \bibinfo{journal}{Phil.\ Mag.}
  \textbf{\bibinfo{volume}{18}}, \bibinfo{pages}{945} (\bibinfo{year}{1968}).

\bibitem{Molokanov89}
  \bibinfo{author}{\bibfnamefont{V.~V.} \bibnamefont{Molokanov}},
  \bibinfo{author}{\bibfnamefont{V.~N.} \bibnamefont{Chebotnikov}},
  \bibnamefont{and} \bibinfo{author}{\bibfnamefont{Y.~K.}
  \bibnamefont{Kovneristyi}}, \bibinfo{journal}{Inorganic Materials}
  \textbf{\bibinfo{volume}{25}}, \bibinfo{pages}{46}
  (\bibinfo{year}{1989}).

\bibitem{Majzoub01}
  \bibinfo{author}{\bibfnamefont{E.~H.} \bibnamefont{Majzoub}},
  \bibinfo{author}{\bibfnamefont{R.~G.} \bibnamefont{Hennig}},
  \bibnamefont{and} \bibinfo{author}{\bibfnamefont{K.~F.} \bibnamefont{Kelton}},
  \bibinfo{journal}{submitted to Phil.\ Mag.\ A} (\bibinfo{year}{2001}).

\bibitem{Yurko59}
  \bibinfo{author}{\bibfnamefont{G.~A.} \bibnamefont{Yurko}},
  \bibinfo{author}{\bibfnamefont{J.~W.} \bibnamefont{Barton}},
  \bibnamefont{and} \bibinfo{author}{\bibfnamefont{F.~G.} \bibnamefont{Parr}},
  \bibinfo{journal}{Acta Cryst.} \textbf{\bibinfo{volume}{12}},
  \bibinfo{pages}{909} (\bibinfo{year}{1959}).

\bibitem{Mackay94}
  \bibinfo{author}{\bibfnamefont{R.}~\bibnamefont{Mackay}},
  \bibinfo{author}{\bibfnamefont{G.~J.} \bibnamefont{Miller}},
  \bibnamefont{and} \bibinfo{author}{\bibfnamefont{H.~F.}
    \bibnamefont{Franzen}}, \bibinfo{journal}{J.\ Alloys Compd.}
  \textbf{\bibinfo{volume}{204}}, \bibinfo{pages}{109} (\bibinfo{year}{1994}).
  
\bibitem{Hen91a}
  \bibinfo{author}{\bibfnamefont{C.~L.} \bibnamefont{Henley}},
  \bibinfo{journal}{Phys. Rev. B} \textbf{\bibinfo{volume}{43}},
  \bibinfo{pages}{993} (\bibinfo{year}{1991}).

\bibitem{Hennig00}
  \bibinfo{author}{\bibfnamefont{R.~G.} \bibnamefont{Hennig}},
  \bibinfo{author}{\bibfnamefont{E.~H.} \bibnamefont{Majzoub}},
  \bibinfo{author}{\bibfnamefont{A.~E.} \bibnamefont{Carlsson}},
  \bibinfo{author}{\bibfnamefont{K.~F.} \bibnamefont{Kelton}},
  \bibinfo{author}{\bibfnamefont{C.~L.} \bibnamefont{Henley}},
  \bibinfo{author}{\bibfnamefont{W.~B.} \bibnamefont{Yelon}}, \bibnamefont{and}
  \bibinfo{author}{\bibfnamefont{S.}~\bibnamefont{Misture}},
  \bibinfo{journal}{Mat.\ Sci.\ Eng.\ A}
  \textbf{\bibinfo{volume}{294-296}}, \bibinfo{pages}{361}
  (\bibinfo{year}{2000}).
  
\bibitem{HennigThesis} \bibinfo{author}{\bibfnamefont{R.~G.}
    \bibnamefont{Hennig}}, Ph.D. thesis, \bibinfo{school}{Washington
    University in St.\ Louis} (\bibinfo{year}{2000}).

\bibitem{Mih96}
  \bibinfo{author}{\bibfnamefont{M.}~\bibnamefont{Mihalkovi{\v c}}},
  \bibinfo{author}{\bibfnamefont{W.~J.} \bibnamefont{Zhu}},
  \bibinfo{author}{\bibfnamefont{C.~L.} \bibnamefont{Henley}},
  \bibnamefont{and} \bibinfo{author}{\bibfnamefont{R.}~\bibnamefont{Phillips}},
  \bibinfo{journal}{Phys.\ Rev.\ B} \textbf{\bibinfo{volume}{53}},
  \bibinfo{pages}{9021} (\bibinfo{year}{1996}).

\end{thebibliography}
\end{document}